# Hybrid-Vehfog: A Robust Approach for Reliable Dissemination of Critical Messages in Connected Vehicles

Anirudh Paranjothi[1], Urcun Tanik[2], Yuehua Wang[2], Mohammad.S. Khan[3]

[1]School of Computer science, University of Oklahoma, – Norman, Oklahoma, USA.
[2]Department of Computer Science, Texas A&M University – Commerce, Texas, USA
[3]Department of Computing, East Tennessee State University, Johnson City, Tennessee, USA.

*Abstract*— Vehicular Ad-hoc Networks (VANET) enable efficient communication between vehicles with the aim of improving road safety. However, the growing number of vehicles in dense regions and obstacle shadowing regions like Manhattan and other downtown areas leads to frequent disconnection problems resulting in disrupted radio wave propagation between vehicles. To address this issue and to transmit critical messages between vehicles and drones deployed from service vehicles to overcome road incidents and obstacles, we proposed a hybrid technique based on fog computing called Hybrid-Vehfog to disseminate messages in obstacle shadowing regions, and multi-hop technique to disseminate messages in non-obstacle shadowing regions. Our proposed algorithm dynamically adapts to changes in an environment and benefits in efficiency with robust drone deployment capability as needed. Performance of Hybrid-Vehfog is carried out in Network Simulator (NS-2) and Simulation of Urban Mobility (SUMO) simulators. The results showed that Hybrid-Vehfog outperformed Cloud-assisted Message Downlink Dissemination Scheme (CMDS), Cross-Layer Broadcast Protocol (CLBP), PEer-to-Peer protocol for Allocated REsource (PrEPARE), Fog-Named Data Networking (NDN) with mobility, and flooding schemes at all vehicle densities and simulation times.

## I. INTRODUCTION

Vehicular Ad-hoc Networks (VANET) has evolved from the Mobile Ad-hoc Networks (MANET) with distinguished characteristics such as high mobility and dynamically changing topology. The main objective of the VANET is to ensure road safety by reducing the number of accidents, optimizing the traffic flow, etc. Advancements in VANET and Fog computing in recent years have gained significant attention in Intelligent Transport Systems (ITS) in terms of broadcasting messages among connected vehicles in an efficient manner [1]. The messages depend on Dedicated Short Range Communication (DSRC) (IEEE 802.11p standards) to establish communication between the vehicles. In general, DSRC has a set of protocols implemented to create a safe driving environment. It contains a dedicated 5.9 GHz band used for vehicular communication. DSRC has one control channel responsible for sending critical messages like information concerning road accidents, traffic jams, roadblocks, etc. and six service channels responsible for sending non-critical messages like personal messages, etc. to nearby vehicles [2]. In addition to critical and non-critical messages, the vehicles send and receive Basic Safety Message (BSM) every 10 ms with the help of control channel, which includes speed, GPS location, brake status, etc.

Two types of communication are used in VANET: 1) Vehicle to Vehicle communication (V2V), and 2) Vehicle to Infrastructure communication (V2I) [3]. For short distance communication of critical messages, V2V communication is employed since the vehicles can communicate with each other directly. The multi-hop technique is used to transmit the messages among them [4]. Although inexpensively reducing communication overhead, it is not suitable for long-range communication due to the transmission delay. Thus, a reliable solution is to use V2I communication which makes use of roadside infrastructures like Road Side Units (RSUs), base stations, Wi-Max towers, etc., to establish communication between the vehicles.

Cloud computing in VANET is commonly known as Vehicular Cloud Computing (VCC). It is used to handle complex tasks in a connected vehicular environment including offloading large files, minimize traffic congestion, encrypting and decrypting messages, etc. [5,6]. However, the limitations of VCC are: 1) High delay in processing and responding requests, 2) High maintenance cost, and 3) High wireless bandwidth cost. Fog computing emerged as an alternative solution to the cloud computing and have gained immediate attention from the academia and industry due to its dynamic nature in creating, incrementing, and destroying the fog nodes.

Fog computing is also known as edge computing and considered as a new revolutionary way of thinking in wireless networking. It is an extension of cloud computing where computations are performed at the edge of the network [7]. Fog computing offers unique services including location awareness, ultra-low frequency and context information [8]. The fog nodes can be created, deployed and destroyed faster when compared to other traditional techniques. To publicize fog computing, open fog consortium created an operational model, reference architecture, and testbed for researchers [9]. In the meantime, cloud in fog computing used popularly to monitor the performance of the network along with resource sharing, resource allocation, etc. using cloud servers.

In a connected vehicle environment, the timely broadcasting of critical messages allows drivers to become aware of emergency situations such that they will have adequate time to make a suitable decision. However, due to frequent topology changes and the limited transmission range of DSRC, delivering messages to their destination is still challenging within a specific amount of time. Though existing techniques provide a solution to this problem, still there is a need for a feasible and effective solution in vehicle-dense regions like downtown areas in Manhattan. This occurs due to obstacle shadowing often caused by tall buildings, which disrupt radio signal propagation between vehicles. A brief explanation of this problem is illustrated in Section III. The solution to this problem is to establish the fog nodes near shadowed regions that enable broadcast messages to the vehicles located in it. A comprehensive review of existing work and its challenges in terms of dissemination of critical messages are briefly explained in Section- II.

The objective of Hybrid-Vehfog is to provide message delivery to a targeted vehicle in a dense region like Manhattan and other downtown areas where it is not possible to establish a continuous connection between the vehicles for reliable communication. To achieve this goal, a hybrid technique is proposed for the dissemination of critical messages in a connected vehicle environment (Hybrid-Vehfog) which yields a faster and more effective solution for the dissemination of critical messages by reducing jitter and channel access time. Hybrid-Vehfog uses the fog computing technique for disseminating messages in obstacle shadowing regions, whereas the multi-hop technique is utilized in non-obstacle shadowing regions to improve communications resiliency by reducing frequency of message drops while increasing efficient resource utilization.

The rest of this paper is organized as follows: Related works are discussed in Section II. A description of the problem is illustrated in Section III. The proposed solution and algorithm for critical message dissemination is presented in Section IV. Based on the proposed approach, a performance evaluation is presented in Section V. Validation is provided through extensive simulation results in Section VI before concluding the paper in Section VII.

## II. RELATED WORKS

Previous authors used either multi-hop (V2V), various roadside infrastructure (V2I), vehicular cloud, and fog nodes to disseminate messages among vehicles in non-obstacle shadowing regions. Liu *et al*. [6] proposed a cloud-based method to disseminate the message between the vehicles, where gateways provide internet access through cellular interfaces to transmit the message. Furthermore, the authors presented the multi-point message dissemination scheme to reduce the packet delay. However, this approach is not suitable for urban scenarios due to resulting transmission delays that are too frequent or extensive in duration. Although Syfullah *et al*. [10] demonstrated a model which combines DSRC and LTE which established a hybrid RSU to broadcast the messages between the vehicles, it is not suitable for vehicle-dense obstacle shadowing regions in urban environments like Manhattan. Feng *et al*. [11] illustrated a VANET-cellular architecture to disseminate the safety messages across vehicles. The authors considered various factors such as link stability, channel quality, signal strength, etc. to select relay nodes that would transmit messages faster in a connected vehicle environment. However, this approach suffers from heavy packet loss. Zhang *et al*. [12] proposed concurrent transmission based broadcast protocol to transmit the messages in an urban environment. The authors divide the transmission into multiple segments and select the forwarder to broadcast the message packets concurrently to reduce the transmission delay. However, the concurrent transmission may lead to a severe packet loss, since large number of packets are generated and transmitted simultaneously between the vehicles.

Sarkar *et al*.[13] discussed the usage of fog computing techniques with the internet of things. Also, the authors performed a comparison of the traditional cloud computing paradigm with fog computing in terms of the internet of things. Tang *et al*. [14] proposed a hierarchical fog computing model for big data analysis in smart cities. Also, the authors analyzed the case study of a smart pipeline system and constructed a working prototype of it to demonstrate its implementation. Ahamad *et al*. [15] proposed a novel framework (Health fog) using fog computing for health and wellness applications. The Health fog combines the data from different sources with an adequate level of security. Preden *et al*. [16] combined the data design approach with fog computing to perform the computation at the edge of the network. Do *et al*. [17] addressed the issue of joint resource allocation and carbon footprint problem of video streaming with the help of fog computing. Aazam and Huh [18] discussed an efficient resource management for fog computing. Also, the authors analyzed various complexities involved in resource allocation of fog.

In this section, existing critical message dissemination strategies are classified into four categories: 1) Transmission of messages using a multi-hop technique, 2) Transmission of messages based on beaconing, 3) Transmission of messages using the vehicular cloud, 4) Transmission of messages with obstacle shadowing.

2.1 Transmission of messages using a multi-hop technique

In this subsection, we discuss the existing approaches where critical messages are delivered to the targeted vehicle using a multi-hop technique. The vehicle transmits the message to a neighboring vehicle (a vehicle in its communication range) until it reaches the targeted vehicle. Santa *et al*. [19] proposed a protocol which uses peer to peer group based technology for transmission of messages between the vehicles. However, the message propagation area is limited in this approach. Hager *et al*. [20] discussed a delayed multi-hop based protocol for message dissemination. Also, they determined the parameters to forward the broadcasted message. Performance of their approach was evaluated using three different Events of Interests (EOI). Peksen *et al*. [21] proposed a protocol which computes the distance between the sender and recipient based on the speed of the vehicles and channel availability. Based on this information messages are relayed from the sender to a targeted vehicle. Libing *et al*. [22] authors proposed a multi-hop dissemination protocol known as Black-burst and multi-channel based multi-hop Broadcast protocol (BMMB) to disseminate emergency messages to the nearby vehicles. However, it is not suitable for a complicated environment such as Manhattan environment, downtown areas, etc. where traffic density is high. Sanguesa *et al.*[23] proposed a Real-Time Adaptive Dissemination (RTAD) system which selects a message dissemination scheme based on the number of informed vehicles and percentage of messages received by each vehicle. However, it is not suitable for transmitting emergency messages due to the significant amount of time consumed in selecting the broadcasting scheme to transmit the messages. Fogue *et al.* [24] discussed a Cooperative Neighbor Position Verification (CNPF) system to transmit the warning messages to the neighboring vehicles. However, the authors mentioned the increase in a number of vehicles allows the adversary nodes to occupy the best position in the network which degrades system performance.

2.2 Transmission of messages based on beaconing

In this subsection, we discuss existing approaches where the critical messages are disseminated to the targeted vehicle as beacons. Shakeel *et al.* [25] proposed an application for beacon messages based on the User Datagram Protocol (UDP). Also, the authors evaluated beacon message size and transmission frequency and found that delivery delay increases linearly based on the packet size of beacon messages. Peksen *et.al.* [21] discussed relaying beacon messages among individual nodes. The individual nodes can function as an ordinary node or relay node based upon the transmitted message. Allouche *et al.* [26] proposed the Cluster-Based Beacon Dissemination Process (CB-BDP) to provide the location of all vehicles in its range. Based on this location, beacon messages will be delivered to the drivers.

2.3 Transmission of messages using the vehicular cloud

In this subsection, existing approaches are discussed that use a cloud computing technique to transmit the messages between vehicles. Taleb *et al.* [27] proposed a framework for smooth migration of required IP service between a data center and 3GPP mobile network. Olariu *et al.* [28] proposed the notion of vehicular clouds and deployed sensors on vehicles, parking areas and streets to provide computation and a communication resource which potentially bring benefits to resource providers as well. However, the potential structure of the vehicular cloud is not discussed in the proposed framework. Taleb *et al.* [29] discussed the possibility of extending vehicular cloud beyond the data center towards the mobile user. Also, the authors presented the challenges involved in mobile network operators. Eltoweissy *et al.* [30] demonstrated the future of vehicular clouds Also, they discussed the challenges involved in vehicular clouds in terms of privacy and security. Mershad *et al.* [31] demonstrated the idea of the cloud providing information to the vehicles whenever needed. Liu *et al.* [6] proposed a model to transmit messages, especially accident information to neighboring vehicles with the help of a vehicular cloud. Although the authors used a mobile gateway as an interface between the cloud and vehicles to broadcast the message, they did not adequately address the effects of mobile gateways and obstacle shadowing while broadcasting messages to vehicles in an urban environment. Syfullah *et al.* [10] discussed hybrid roadside unit for disseminating the critical messages to the neighboring vehicles with the help of vehicular cloud network, digital content network, infrastructure cloud network and server to a cloud network. However, this approach is not suitable for urban scenarios due to the various delays associated with the transmission of messages. Also, it is not suitable in obstacle shadowing regions.

2.4 Transmission of messages with obstacle shadowing

Sommer *et al.* [32] proposed a simulation model to estimate the effect of obstacles on radio communication between vehicles. In particular, the authors presented a model to estimate signal attenuation and path loss caused by the obstacles. Carpenter [33] proposed a model to validate the obstacle shadowing and presented the accuracy of a deterministic fading model in estimating the performance of VANET safety applications.

Most of the approaches discussed above are using a multi-hop technique or beaconing or the vehicular cloud to disseminate critical messages to near-by vehicles. However, limitations of the [19-21] techniques include, a high packet error rate, a high transmission delay, the retransmission of messages and the shortcomings of [25,26] involve routing overhead and a high packet loss rate.

III. PROBLEM DESCRIPTION

Radio transmissions are heavily affected by shadowing effects commonly known as obstacle shadowing. Finding a solution for this problem plays an important role in establishing communication between vehicles in urban environments where buildings block radio propagation, as represented in Fig. 1. Assume vehicle V1 is the sender that needs to broadcast critical messages to nearby vehicles (receivers) V2, V3, and V4.

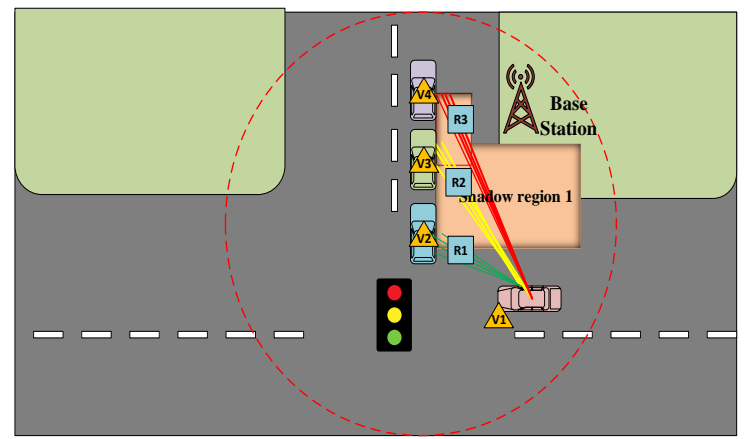

**Fig. 1: Obstacle Shadowing**

To provide a solution, we divided this problem into three zones to denote regions (R1, R2, and R3) as represented in the form of green, yellow and red lines respectively. Here, the nearby vehicles V2, V3, and V4 are located in the transmission range of a base station associated with a sender. The vehicle V2 is in R1 where the message can be sent directly using a hopping technique, vehicle V4 is situated in R3 and its radio transmissions are blocked by shadowing in the same region. It leads to a situation where the message is getting dropped in the middle without reaching the destination. The vehicle V3 is in region R2 where the message may be sent directly or may be dropped without reaching the destination (uncertain region) which increase the complexity of the system. To simplify and increase the probability of message delivery we combined the regions R2 and R3 into a single region R2 as shown in Fig. 1, to overcome the shadowing effects caused by obstacles like tall buildings in a Manhattan and other downtown regions, we developed a hybrid technique for the successful dissemination of critical messages reliably under these conditions. A detailed explanation of our proposed approach is illustrated in next section.

## IV. PROPOSED SOLUTION – HYBRID VEHFOG

In a dense urban environment, it is difficult for vehicles in close proximity to reliably establish continuous communication between them due to obstacle shadowing delays and drops caused by intervening tall buildings. To promote continuously reliable communication between the vehicles we developed a hybrid architecture where the critical messages are delivered to the nearby vehicles within the transmission range of a base station sending messages either by using a multi-hop technique or the fog computing, as needed. In our approach, we concentrated only on the vehicles in the transmission region of a base station associated with a sender. Fig. 2 represents the proposed architecture for the dissemination of critical messages, in which dissemination of critical messages using the fog computing is illustrated in Case 1 and dissemination of critical messages broadcasting using a multi-hop technique is illustrated in Case 2.

**Case 1: Dissemination of critical messages using fog computing**

In a connected vehicular environment such as VANET, vehicles are highly connected to each other at all times based on V2V and V2I techniques. But, when vehicles encounter shadowing regions in dense urban environments, critical messages transmitted among vehicles can be dropped due to intermittent connections resulting from obstacle shadowing from obstacles such as tall buildings. In such cases, fog computing a crucial role in disseminating messages.

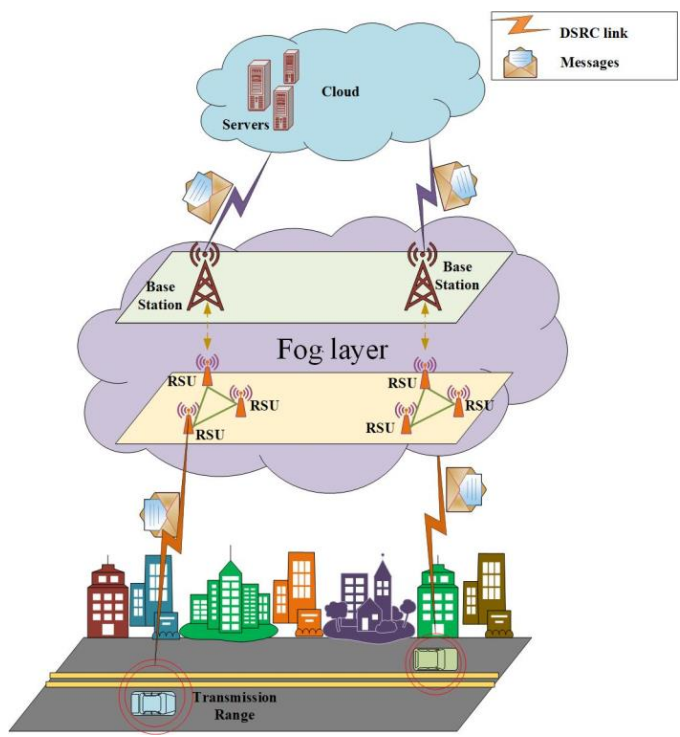

(a)

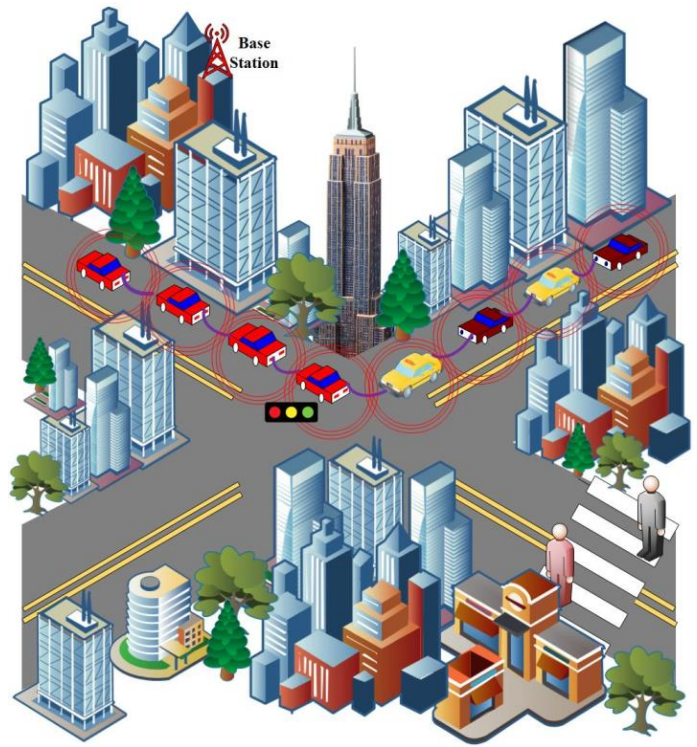

(b)

**Fig. 2: Dissemination of messages using Hybrid-Vehfog a) Fog computing technique and b) Multi-hop technique**

Fog layer is located at the edge of a network. It consists of fog nodes, which includes access points, gateways, RSUs, base station, etc. In our approach, RSUs and base stations play a major role in disseminating the messages. Fog layer can be static at a fixed location or mobile on moving carriers such as in the vehicular environment. They are responsible for processing the information received from the vehicles and temporarily store it or broadcast over the network. It can be used widely for latency-sensitive applications like broadcasting emergency messages, etc. Cloud in fog computing is used to keep track of the resources allocated to each fog node and to manage interaction and interconnection among workloads on a fog layer, popularly known as fog orchestration.

As the vehicles are aware of their locations in relation to the base station, the system deploys and broadcasts the critical messages to the fog layer and when it encounters the obstacle shadowing region. As a result, the messages are disseminated to the vehicles in the shadowing region seamlessly through the fog nodes.

**Case 2: Dissemination of critical messages using multi-hop technique**

Consider the same situation discussed in Case 1 whereas the vehicles can communicate with each other directly using a multi-hop technique which means the vehicles are in non-shadowed regions, allowing communication to be established directly between vehicles. The main advantage of this approach is that vehicles are able to communicate with each other directly without any external technique such as fog computing In this approach, an On-Board Unit (OBU) is used to establish multi-hop communication between the vehicles. When a new vehicle enters the region, critical messages, such as hazard alerts, can be delivered to the vehicle based on a multi-hop technique or the fog nodes based on its location.

4.1. Analysis of Hybrid-Vehfog

In this analysis, we calculated the power at a receiver end. Analogous to the approaches [32,33], we thus conceive our model to be a generic expansion of a well-established shadowing model. In general, it is expressed in the form of Eqn. (1) [32].

$$P_r = P_t + G_t + G_r - \sum L_x \qquad (1)$$

Such that $P_r$ is the received power, $P_t$ is the transmitted power, $G_t$ is the antenna gain at the transmitter end, $G_r$ is the antenna gain at the receiver end and $L_x$ is the loss of effect during transmission. In our system, the major transmission loss is due to obstacle shadowing, as formulated in next sub section.

4.2. Obstacle modelling

Obstacle modeling is formulated based on our problem description and proposed solution (Case 1 and Case 2). Assume the transmission range of a vehicle ($T_{base}$) is in the form of a circle and divided into two regions such as R$_1$, and R$_2$, where R$_1$ is the non-obstacle shadowed region, and R$_2$ is the obstacle shadowed region ($O_{shadow}$. The power levels in the obstacle shadowed regions are measured in decibels (dB) but, to calculate the area of regions R$_1$ and R$_2$ we need to express the $O_{shadow}$ in meters. It can be done by:

$$d = \frac{10(O_{shadow} - 32.44 - 20\log(f))}{20} * 1000 \qquad (2)$$

Where $d$ is the distance, and $f$ is the frequency. For our approach, $f$ = 5.9 GHz.

The transmission range of a vehicle ($T_{base}$) and area of the zones demarcated as regions (R$_1$ and R$_2$) are calculated as follows: 1) $T_{base} = \pi r^2$, 2) $R_1 = T_{base} - \pi d^2$, and 3) $R_2 = \pi d^2$. Where, $r$ is the radius, represented in meters (m).

For each obstacle in the line of sight between the vehicles, represented in Fig. 1 and Fig. 2(a), the effect of obstacle shadowing ($O_{shadow}$) region is calculated as follows:

$$O_{shadow} = \alpha n + \beta l_{obs} \qquad (3)$$

Where $n$ is the number of times an obstacle encountered, $l_{obs}$ is the total length of an obstacle, $\alpha$ represents the attenuation due to the exterior wall, and $\beta$ represents the approximate internal structure of an obstacle. According to [32], the generic equation to calculate the power received at a receiver end due to obstacle shadowing is formulated as follows:

$$P_r = P_t + G_t + G_r - O_{shadow} \qquad (4)$$

Where $P_r$ is the received power, $P_t$ is the transmitted power, $G_t$ is the antenna gain at the transmitter end, and $G_r$ is the antenna gain at the receiver end. Based on the transmitted power and received power in Eqn. (4), we can determine the obstacle shadowing in dense urban environments with high vehicle densities.

4.3. Delay Analysis

Delay refers to the time taken for a packet to be transmitted across a network from source to destination. It is an additive metric, and thus, overall delay (end-to-end delay) equal to the sum of delays in each hop during a multi-hop data transmission (Case 1 and Case 2). According to [34], the single-hop delay of a network ($D$) can be calculated as follows:

$$D = t_{trans} + t_q + t_{cont} + t_{proc} + t_{prop} \qquad (5)$$

Where $t_{trans}$ is the transmission delay, $t_q$ is the queuing delay, $t_{cont}$ is the contention delay, $t_{proc}$ is the processing delay, and $t_{prop}$ is the propagation delay.

## 4.4. Message success rate analysis

Message success rate directly impacts the performance of the system. Thus, an increase in message success rate improves the performance of Hybrid-Vehfog. It is calculated as follows:

$$M_{success} = \frac{P_{msg} * D}{N_{users}} \quad (6)$$

$$\begin{cases} 0.5 \leq M_{success} \leq 1, \text{ message disseminated using} \\ \qquad\qquad\qquad\qquad \text{multi hop} \\ 0 \leq M_{success} < 0.5, \text{ message disseminated using} \\ \qquad\qquad\qquad\qquad \text{fog computing} \end{cases}$$

Where $M_{sucess}$ is the message success rate, $P_{msg}$ is the probability of message delivery, and $N_{users}$ is the number of users associated with the system.

From Eqn. (6), we can observe that Hybrid-Vehfog provides guaranteed message delivery to the nearby vehicles in an urban environment using the fog computing or multi-hop techniques (Case 1 and Case 2). As a result, the robustness of our system is relatively high when compared with previous protocols.

## 4.5. Algorithm

___
**Algorithm** Hybrid-Vehfog (input_msg)
___
1. **scan** trans_range ($V_x$)
2. **calculate** $n$
3. **if** ($n > 0$) **then**
4.   **for**($i=1;i<=n;i++$)
5.     $loc[i]$= obstacle_shadowing($i$)
6.     **if**($loc[i] == 0$) **then**
7.       **call** multi_hop_tech(input_msg)
8.       **print message sent using multi-hop technique**
9.     **else if** ($loc[i] == 1$) **then**
10.       **establish** fog_layer(input_msg)
11.       **print message sent using fog technique**
12.     **endif**
13.   **endfor**
14. **else**
15.   **print no nearby vehicles were located**
16. **endif**
17. **if** ($V_y == 1$) **then**
18.   **repeat** steps 1 to 16
19. **endif**

The proposed algorithm for broadcasting critical messages takes the input of a critical message *(input_msg)* from a vehicle and broadcasts the message to nearby vehicles. This algorithm works as follows: first, the set of neighboring vehicles in the transmission range of a base station associated with a sender is calculated. Then, *trans_range ($V_x$)* is used to discover the number of vehicles that are in the range of the base station. In the event the number of number of vehicles is greater than zero, the location of vehicles is determined using the *obstacle_shadowing()* function. This function returns the binary value 0 or 1 to the *loc* variable. The value 0 indicates a vehicle is located in a non-shadowed region so the message can be broadcasted using the multi-hop technique. The corresponding *multi_hop_tech(msg)* is used to send the message using the multi-hop technique. The value 1 represents the vehicles is in a shadowed region and hence the messages are broadcasted using the fog computing technique (i.e., *fog_layer(input_msg)*.

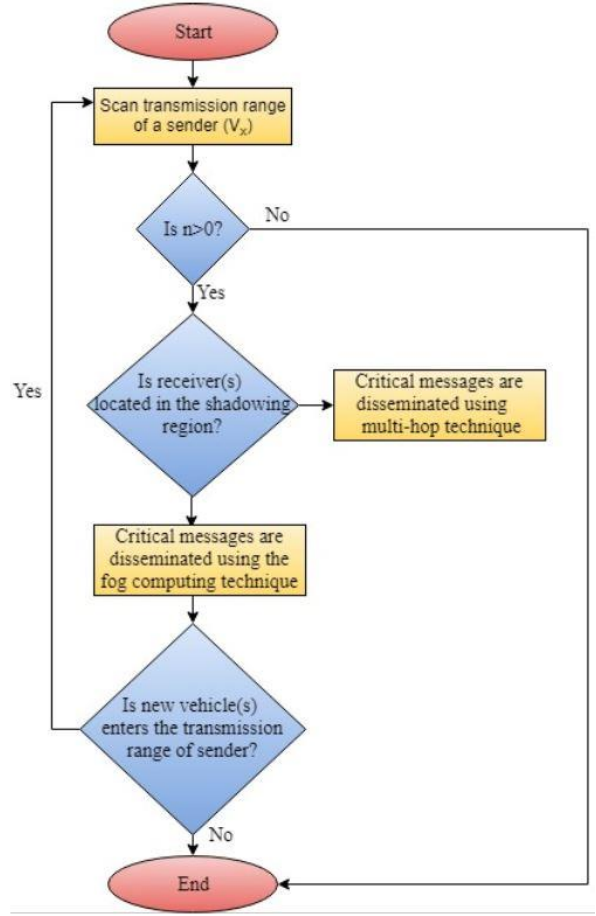

**Fig. 3: Flowchart of Hybrid-Vehfog**

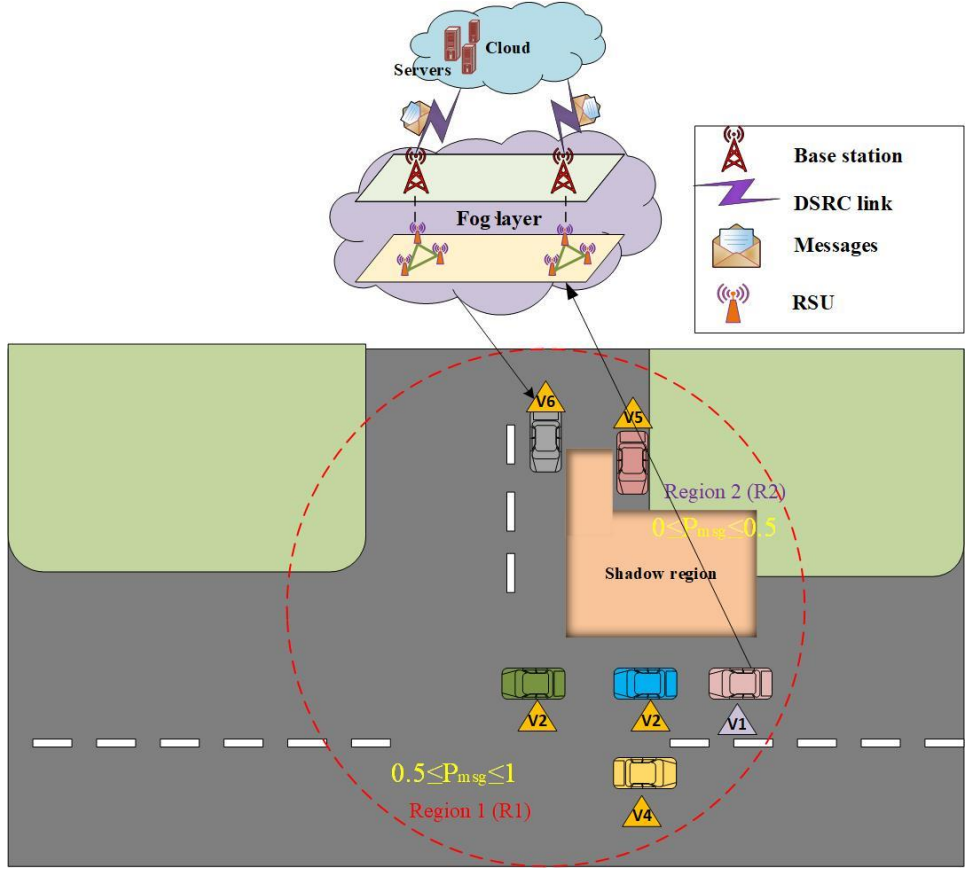

**Fig. 4: Example Scenario**

**Table 1: Notations used in Hybrid-Vehfog**

| Variables and Functions | Purpose |
|---|---|
| $V_x$ | Vehicle broadcasts the message (sender) |
| $n$ | Number of vehicles in the transmission range of a sender (i.e., receiver(s)) |
| $loc$ | Either 0 or 1, based on this message the dissemination technique is determined |
| $V_y$ | New vehicle enters the transmission range of base station associated with a sender |
| trans_range() | Calculates the number of near-by vehicles located in transmission range of sender |
| obstacle_shadowing() | Determines the location of the intended recipients (receiver and returns the value to the loc variable) |
| multi_hop_tech() | Used to transmit the message using multi-hop technique |
| fog_layer() | Used to transmit the message using fog computing technique |

If a new vehicle ($V_y$) enters the transmission range of a base station associated with a sender ($V_x$), the whole process described above is repeated until the message gets delivered. A flowchart representation of this algorithm is presented in Fig. 3 and notations used in this algorithm are presented in Table 1.

When a new vehicle enters the location (*if ($V_y == 1$)*) then steps 1 to 3 will be repeated to identify if the new vehicle is in the transmission range of the vehicle that broadcasts the critical message. Based on this result steps 4 to 10 will be repeated to determine the technique used to deliver the message (the message will be delivered using the fog computing technique or multi-hop technique).

V. PERFORMANCE EVALUATION

The simulation of our algorithm is performed using ns-2 and SUMO simulators. We considered the following metrics to measure the performance of our approach:

- End-to-End delay: Time taken for a packet to be transmitted across a network from source to destination

- Collision ratio: The number of packets colliding across a network before reaching the destination

- Probability of message delivery: Probability of the message is delivered to the receiver

5.1 Simulation scenario

Consider the situation where vehicle V1 needs to broadcast a critical message to the nearby vehicles within its transmission range. First, it will search for the vehicles in region 1 (R1) where the message can be transmitted directly using a hopping technique. In this scenario vehicles V2, V3, and V4 are located in R1 so the message is transmitted directly using the multi-hop technique, which is represented in Fig. 4. Second, the sender (V1) will search for the vehicles in R2 where the message cannot be transmitted directly due to obstacle shadowing and thus, a fog layer is created near the sender (V1) to broadcast the message. In this scenario, V5 and V6 are the vehicles located in the obstacle shadowed region. Hence, the message is delivered using the fog computing technique. Cloud is used to keep track of the resources allocated to each fog node

5.2. Simulation Setup

Simulation of our algorithm is carried out using two simulators as discussed above. For the movement trace of vehicles, we used open source traffic simulator SUMO to generate the flow of vehicles and to compare our algorithm with existing approaches the ns-2 simulator is used. The simulations were performed in two ways; First, we compared the performance of our fog approach with PEer-to-Peer protocol for Allocated REsource (PrEPARE) and fog-Named Data Networking (NDN) with mobility. In the second approach, we compared the performance of our proposed hybrid approach (Hybrid-Vehfog) with the Cloud-assisted Message Downlink Dissemination Scheme (CMDS), Cross-Layer Broadcast Protocol (CLBP) and flooding schemes. Specifically, we assume that critical messages are broadcasted only to vehicles located in the transmission range of a base station associated with a sender. A brief explanation of these existing approaches is discussed below:

- *PrEPARE:* In this approach, the messages are disseminated to the nearby vehicles with the help of a collaborative network [35]. We compared our approach with PrEPARE since it dealt with broadcasting messages to nearby vehicles located in the range of sender.

- *Fog-NDN with mobility*: In Fog-NDN with mobility, the authors illustrated the possibility of disseminating critical messages to neighboring vehicles using fog computing technique [10]. Hence, we compared it with our fog approach.

- *CMDS*: We compared our approach with CMDS since it dealt with broadcasting messages, especially accident information, to nearby vehicles with the help of mobile gateways [6].

- *CLBP*: In CLBP, using a relay selection scheme the emergency messages are broadcasted to the vehicles involved in inter-vehicle communication using a multi-hop technique [36]. Since the multi-hop technique is used in message broadcasting, we compared the performance of our system with this approach.

- *Flooding*: In this approach, flooding starts with a source node that transmits the message to the neighboring vehicles. The neighbors who received messages again retransmit the message to their neighbors. This process continues until all vehicles in a network receive the message. Our approach is comparable since flooding provides guaranteed message delivery.

The simulations were carried out based on the following parameters, represented in Table 2.

**Table 2: Parameters used in simulation**

| Parameters | Value |
| --- | --- |
| Road length | 10 km |
| Number of vehicles/nodes | 50-300 |
| Number of lanes | 3 |
| Vehicle speed | 30-50mph |
| Transmission range | 300m |
| Critical message size | 256bytes |
| Simulator used | ns-2, SUMO |
| Data rate | 2Mbit/s |
| Technique used | Multi-hop, fog computing |
| Protocol | IEEE802.11p |
| CW Min/Max | 31/1023 |

VI. SIMULATION RESULTS

6.1. Comparison of our fog approach with other protocols

As mentioned above, we compared our fog computing approach with PrEPARE and fog-NDN with mobility. The metrics we considered for simulations are 1) End-to-end delay, 2) Collision ratio and 3) Probability of message delivery. The probability of message delivery of our fog approach was observed to be higher high due to the location awareness with the help of a base station. Hence, it provides the guaranteed message to the vehicles situated in obstacle shadowing region. Whereas in PrEPARE and Fog-NDN with mobility, a message drop is likely during transmission. In addition, the probability of message delivery is low as the number of users increases, which affects the system load, represented in Fig. 5.

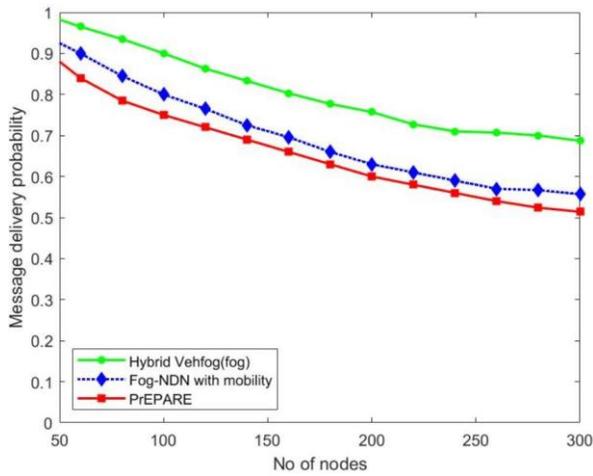

**Fig. 5: Probability of message delivery using fog computing**

End-to-end delay of PrEPARE and fog-NDN with mobility was observed to be higher due to the various delays associated with message transmission. But in our fog approach, knowledge of nearby vehicles including the position significantly reduces the route setup time and propagation time across a network. Hence, it delivers the message much faster when compared to other protocols. The end-to-end delay increases when the number of users increases in a system due to numerous packets that need to be transmitted at a given time, represented in Fig. 6.

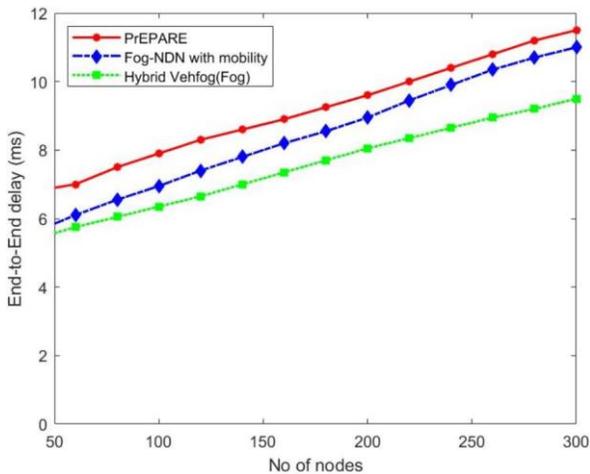

**Fig. 6: End-to-End delay using fog computing**

The collision ratio of our fog computing approach was observed to be lower due to the number of packets (i.e, critical messages) delivered to the nearby vehicles at a given time. This is because our fog approach disseminates critical messages to the vehicles situated in the obstacle shadowing region. But PrEPARE and fog-NDN with mobility rely upon mobile nodes including fog for transmission of messages which results in a packet collision, represented in Fig. 7.

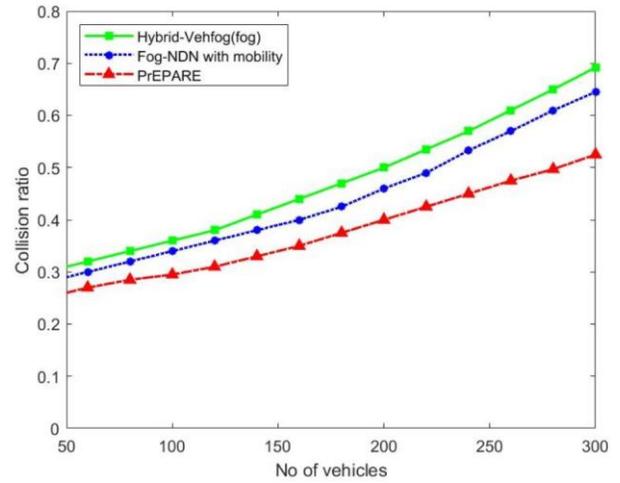

**Fig. 7: Collision ratio using fog computing**

6.2. Comparison of our hybrid fog approach with other protocols

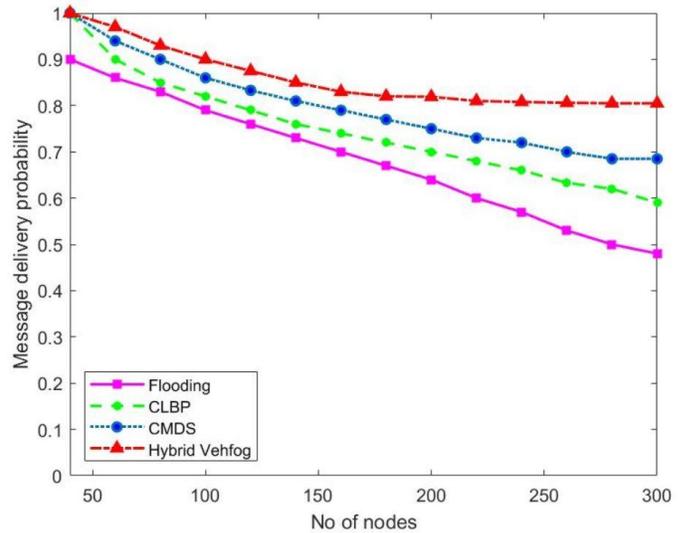

**Fig. 8: Probability of message delivery**

We compared the performance of our algorithm in terms of the end-to-end delay, collision ratio and probability of message delivery with CLBP, CMDS, and flooding protocols. The probability of message delivery using other protocols is relatively low when compared to our approach, as represented in Fig. 8. In CLBP and flooding the messages are disseminated using a multi-hop technique which makes it more likely that a message is dropped in the obstacle shadowed regions. The messages are transmitted using mobile gateways in the CMDS protocol, but mobile gateways are used in transmitting critical messages between a vehicle and the cloud. As a result, this may lead to a message failure situation. In our approach, the messages are transmitted to the vehicles with the help of a fog layer in shadowed regions which ensures guaranteed message delivery and thus, it outperforms other protocols by increasing the probability of message delivery.

A comparison of the end-to-end delay of our approach with other schemes is presented in Fig. 9. The results showed that the end-to-end delay of our approach is lower than that of the CLBP, CMDS, and flooding algorithms. In our proposed approach, messages are disseminated to other vehicles with the help of a base stations and RSUs in the fog layer. The base station is aware of the location of all vehicles situated in its transmission range which helps in reducing the time taken for an initial setup across a network from source to destination and thus, the end-to-end delay of the Hybrid-Vehfog is relatively lower than other protocols.

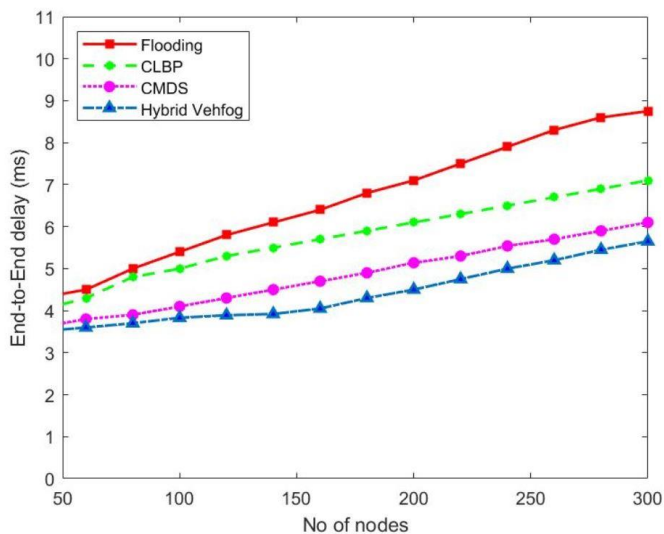

**Fig.9: End-to-End delay**

In order to observe the number of packets that were dropped without reaching their destination, we broadcasted the critical messages to nearby vehicles at a time interval (t1).

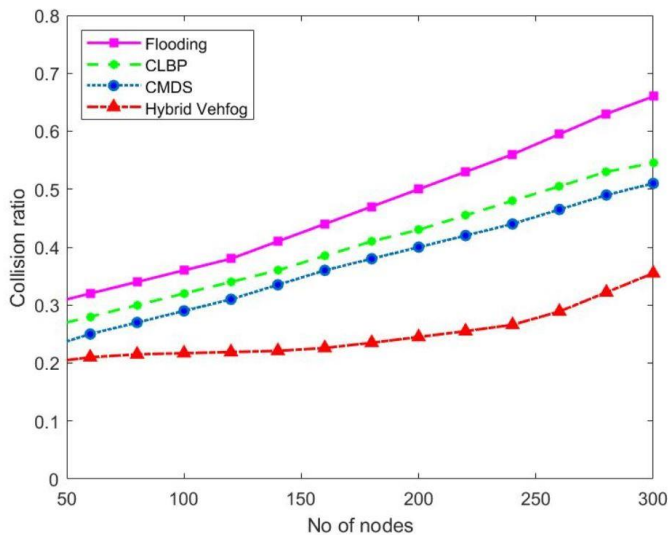

**Fig. 10: Collision ratio**

The collision ratio of our approach was observed to be lower than that of the CLBP, CMDS, and flooding protocols. Our approach provides guaranteed message delivery to the targeted vehicles whereas in other schemes there is a high chance of message transmission failure, a situation which leads to the retransmission of input messages. Accordingly, the number of packets generated in a time interval (t1) increases, which in turn increases the collision ratio, as represented in Fig. 10.

## VII. CONCLUSION AND FUTURE WORK

We studied the vehicle disconnection problem frequently encountered in a connected convoy, notably in a vehicle dense Manhattan-like urban environment and other downtown areas in which transmitted messages may be lost due to obstacle shadowing. In this paper, we developed a hybrid technique based on fog computing called Hybrid-Vehfog to disseminate messages in obstacle shadowing regions, and multi-hop technique to disseminate messages in non-obstacle shadowing regions. To evaluate the performance of our proposed approach, extensive simulation was performed. The results indicated that Hybrid-Vehfog is a robust and efficient approach to reliable communication which provides the best performance when compared with PrEPARE, Fog-NDN with mobility, CLBP, CMDS and flooding algorithms. This encourages us to implement the proposed Hybrid-Vehfog in a real work-in-progress testbed with real-world ground mobile nodes, unmanned aerial vehicles, and various network connections between them to further validate its effectiveness and high performance. We also plan to investigate computer vision-assisted environmental sensing and reliable dissemination of critical messages. Future work is planned with visual data collection, enabling the fog nodes to estimate and predict vehicle telemetry utilizing real-time optical flow measurements, which help make better decisions on message dissemination.

## REFERENCES

[1] S. Khan, L. Senlin, W. Chao, P. Limin, and C. Qianrou, "PIaas: Cloud-oriented secure and privacy-conscious parking information as service using VANET," *Computer Networks,* vol. 124, no. 1, pp. 33-45, 2017.

[2] K. C. Dey, A. Rayamajhi, M. Chowdhury, P. Bhavsar, and J. Martin, "Vehicle-to-vehicle (V2V) and vehicle-to-infrastructure (V2I) communication in a heterogeneous wireless network–Performance evaluation," *Transportation Research Part C: Emerging Technologies,* vol. 68, no 2, pp. 168-184, 2016.

[3] A. Paranjothi, M. S. Khan, M. Nijim, and R. Challoo, "MAvanet: Message authentication in VANET using social networks," *IEEE 7th Annual Ubiquitous Computing, Electronics & Mobile Communication Conference (UEMCON)*, pp. 1-8, 2016.

[4] Z. Cao, Q. Li, H. W. Lim, and J. Zhang, "A multi-hop reputation announcement scheme for VANET," *Proceedings of IEEE International Conference on Service Operations and Logistics, and Informatics*, pp. 238-243, 2014.

[5] A. Paranjothi, M. S. Khan, and M. Nijim, "Survey on Three Components of Mobile Cloud Computing: Offloading, Distribution and Privacy," *Journal of Computer and Communications,* vol. 5, no. 06, pp. 1-31, 2017.

[6] B. Liu, D. Jia, J. Wang, K. Lu, and L. Wu, "Cloud-assisted safety message dissemination in the VANET-cellular heterogeneous wireless network," *IEEE Systems Journal*, vol. 11, no. 1, pp. 128-139, 2017.

[7] A. Paranjothi, M. S. Khan, and M. Atiquzzaman, "DFCV: A Novel Approach for Message Dissemination in Connected Vehicles Using Dynamic Fog," *6th IFIP International Conference on Wired/Wireless Internet Communications,* vol. 5, no. 06, pp. 1-12, 2018.

[8] O. Salman, I. Elhajj, A. Kayssi, and A. Chehab, "Edge computing enabling the Internet of Things," IEEE 2nd World Forum on Internet of Things (WF-IoT), pp. 603-608, 2015.

[9] Open fog consortium, Available: https://www.openfogconsortium.org/. [Accessed: Jan.28, 2019].

[10] M. Syfullah, and J. M. Y. Lim, "Data broadcasting on Cloud-VANET for IEEE 802.11p and LTE hybrid VANET architectures," *3rd International Conference on Computational Intelligence & Communication Technology (CICT)*, pp. 1-6, 2017.

[11] D. Feng, M. Yajie, Z. Xiaomao, and H. Kai, "A Safety Message Broadcast Strategy in Hybrid Vehicular Network Environment," *The Computer Journal,* vol. 61, no. 6, pp. 789-797, 2017.

[12] X. Zhang, L. Yan, H. Zhang, and D. K. Sung, "A Concurrent Transmission Based Broadcast Scheme for Urban VANET," *IEEE Transactions on Mobile Computing*, 2018.

[13] S. Sarkar, S. Chatterjee, and S. Misra, "Assessment of the Suitability of Fog Computing in the Context of Internet of Things," *IEEE Transactions on Cloud Computing*, vol.1, no. 99, pp.1-14, 2018.

[14] B. Tang, Z. Chen, G. Hefferman, T. Wei, H. He, and Q. Yang, "A Hierarchical Distributed Fog Computing Architecture for Big Data Analysis in Smart Cities," *Proceedings of the ASE BigData & Social Informatics*, pp. 1-28, 2015.

[15] M. Ahmad, M. Bilal Amin, S. Hussain, B. Kang, T. Cheong, and S. Lee. "Health Fog: A novel framework for health and wellness applications," *Journal of Supercomputing*, vol. 72, no. 10, pp.3677-3695, 2016.

[16] J. Preden, J. Kaugerand, E. Suurjaak, S. Astapov, L. Motus, and R. Pahtma, "Data to decision: pushing situational information needs to the edge of the network," *IEEE International Multi-Disciplinary Conference on Cognitive Methods in Situation Awareness and Decision*, pp. 158-164, 2015.

[17] C. T. Do, N. H. Tran, Chuan Pham, M. G. R. Alam, Jae Hyeok Son, and C. S. Hong, "A proximal algorithm for joint resource allocation and minimizing carbon footprint in geo-distributed fog computing," *International Conference on Information Networking (ICOIN)*, pp. 324-329, 2015.

[18] M. Aazam and E. N. Huh, "Dynamic resource provisioning through Fog micro datacenter," *IEEE International Conference on Pervasive Computing and Communication Workshops (PerCom Workshops),* pp. 105-110, 2015.

[19] J. Santa, A. F. Gómez-Skarmeta, and M. Sánchez-Artigas, "Architecture and evaluation of a unified V2V and V2I communication system based on cellular networks," *Computer Communications,* vol. 31, no. 12, pp. 2850-2861, 2008.

[20] M. Hager, L. Wernecke, C. Schneider, and J. Seitz, "Vehicular Ad Hoc networks: Multi-hop information dissemination in an urban scenario," *38th International Conference on Telecommunications and Signal Processing (TCP)*, pp. 65-70, 2015.

[21] Y. Pekşen, and T. Acarman, "Multihop safety message broadcasting in VANET: A distributed medium access mechanism with a relaying metric," *International Symposium on Wireless Communication Systems (ISWCS),* pp. 346-350, 2012.

[22] W. Libing, N. Lei, F. Jing, H. Yanxiang, L. Qin, and W. Dan, "An Efficient Multi-hop Broadcast Protocol for Emergency Messages Dissemination in VANET", *Chinese Journal of Electronics*, vol. 26, no. 3 pp. 614-623, 2017.

[23] J. A. Sanguesa, M. Fogue, P. Garrido, F. J. Martinez, J. Cano, C. T. Calafate, and P. Manzoni, "RTAD: A real-time adaptive dissemination system for VANET", *In Computer Communications*, vol. 60, pp. 53-70, 2015.

[24] M. Fogue, F.J. Martinez, P. Garrido, M. Fiore, C.F. Chiasserini, C. Casetti, J.C. Cano, C.T. Calafate, and P. Manzoni, "Securing Warning Message Dissemination in VANET Using Cooperative Neighbor Position Verification," in *IEEE Transactions on Vehicular Technology*, vol. 64, no. 6, pp. 2538-2550, 2015.

[25] S. M. Shakeel, M. Ould-Khaoua, O. Rehman, A. Maashri, and H. Bourdoucen, "Experimental Evaluation of Safety Beacons Dissemination in VANET", *Procedia Computer Science*, vol. 56, no. 1, pp. 618-623, 2015.

[26] Y. Allouche, and M. Segal, "Cluster-Based Beaconing Process for VANET", *Vehicular Communications*, vol. 2, no. 2, pp. 80-94, 2015.

[27] T. Taleb, and A. Ksentini, "Follow me cloud: Interworking federated clouds & distributed mobile networks," *IEEE Networks*, vol. 27, no. 5, pp. 12–19, 2013.

[28] S. Olariu, T. Hristov, and G. Yan, "The next paradigm shift: From vehicular networks to vehicular clouds," in *Mobile Ad Hoc Networking: The Cutting Edge Directions*, vol. 19, no. 1, pp. 645–700, 2013.

[29] T. Taleb, "Towards carrier cloud: Potential, challenges, & solutions," *IEEE Wireless Communications.*, vol. 21, no. 3, pp. 80–91, Jun. 2014.

[30] M. Eltoweissy, S. Olariu, and M. Younis, "Towards autonomous vehicular clouds," *Proceedings of Ad Hoc Networks*, vol. 49, pp. 1–16, 2010.

[31] K. Mershad, and H. Artail, "A framework for implementing mobile cloud services in VANET," *Proceedings of IEEE 6th International Conference Cloud Computing*, pp. 83–90, 2013.

[32] C. Sommer, D. Eckhoff, R. German and F. Dressler, "A computationally inexpensive empirical model of IEEE 802.11p radio shadowing in urban environments," *8th International Conference on Wireless On-Demand Network Systems and Services*, pp. 84-90, 2011.

[33] S. E. Carpenter, "Obstacle Shadowing Influences in VANET Safety," *IEEE 22nd International Conference on Network Protocols*, pp. 480-482, 2014.

[34] K. L. K. Sudheera, M. Ma, G. G. M. N. Ali, and P. H. J. Chong, "Delay efficient software-defined a networking based architecture for vehicular networks," *Proceedings of 15th International Conference on Communication Systems (ICCS)*, pp. 1-6, 2016.

[35] R.I. Meneguette, and A. Boukerche. "Peer-to-peer protocol for allocated resources in vehicular cloud based on V2V communication." *Wireless Communications and Networking Conference (WCNC),* pp. 1-6. 2017.

[36] Y. Bi, L. X. Cai, and H. Zhao, "A Cross-Layer Broadcast Protocol for Multi-hop Emergency Message Dissemination in Inter-Vehicle Communication," *Proceedings of IEEE International Conference on Communications (ICC)*, pp. 1-5, 2010.